\documentstyle[prb,aps,multicol,epsfig]{revtex}

\begin{document} \bibliographystyle{unsrt}
\input{epsf}
\title{Dynamics of the diluted Ising antiferromagnet
Fe$_{0.31}$Zn$_{0.69}$F$_{2}$}
\author
{
K. Jonason$^{a}$, P. Nordblad $^{a}$ and F. C. Montenegro$^{b}$
}
\address
{
$^a$Dept. of Materials Science, Uppsala University, Box 534, S-751 21 Uppsala,
Sweden\\
$^b$Departamento de Fisica, Universidade Federal de Pernambuco, 50739-901
Recife, PE, Brasil
}

\maketitle

\begin{abstract}
The diluted Ising antiferromagnet, Fe$_{0.31}$Zn$_{0.69}$F$_{2}$, has 
been investigated by dynamic susceptibility measurements in zero and 
and finite applied dc-fields.  In zero field, there is a para- to 
antiferromagnetic phase transition at $T_{N}$ $\approx {20.0 K}$, 
followed by a dramatic slowing down of the dynamics at low 
temperatures.  The latter described by a pure Arrhenius law.  The 
effect of a superposed dc-field on the antiferromagnetic phase 
transition and on the low-temperature dynamics is investigated, and a 
comprehensive static and dynamic phase diagram in the $H-T$ plane is 
derived.  In agreement with earlier results on the same system, 
$T_{N}(H)$follows a random-exchange Ising model to random-field Ising 
model crossover scaling for fields $H \leq$1.5 T. A random-field 
induced glassy dynamics appears for higher values of H, where the 
antiferromagnetic phase transition is destroyed.  The low-temperature 
dynamics shows striking similarities with the behavior observed in 
reentrant antiferromagnets.
\end{abstract}

\vskip 0.2cm

\centerline{PACS numbers: 75.40.Gb 75.50.Lk}
\vskip 0.2cm

\bigskip
\begin{multicols}{2}
\narrowtext

\section{introduction}

The diluted Ising antiferromagnet Fe$_{x}$Zn$_{1-x}$F$_{2}$ in an
external magnetic field has proven to be a good model system for the
random-exchange Ising model (REIM) ($H\approx$ 0) and the random field
(RFIM) Ising model ($H>$ 0) \cite{one,two}.  In the original
derivation of the equivalence between the RFIM and a diluted Ising
antiferromagnet in a uniform applied field, weak dilution and small
values of $H/J$ were assumed \cite{three} ($J$ is the magnitude of the
exchange interaction).  In the Fe$_{x}$Zn$_{1-x}$F$_{2}$ system, weak
dilution implies Fe concentrations well above the percolation
threshold \cite{four} $x_{p}$=0.25 and the most convincing
experimental results on the RFIM critical behaviour have been obtained on
samples with  \cite{one,two} $x\geq$0.46.  On the other hand,
interesting dynamic properties may become observable in the limit of
strong dilution.  RFIM systems have been argued to attain extremely
long relaxation times at temperatures near $T_{c}$ \cite{five} and for
large values of $H/J$, where the ordered phase is destroyed, it has
been argued that a glassy phase will appear, even without exchange
frustration being present in the system \cite{six}.  Experimental
results on Fe$_{x}$Zn$_{1-x}$F$_{2}$ samples of concentration at or
slightly above $x_{p}$ have revealed some dynamic properties similar
to those of conventional spin glasses \cite{seven,eight,nine,ten}.  In
a recent paper \cite{eleven} it was found that the percolating
threshold sample Fe$_{0.25}$Zn$_{0.75}$F$_{2}$ exhibits magnetic
ageing, a typical spin glass feature, whereas the slowing down of the
dynamics followed an Arrhenius law, i.e.  it did not support the
existence of a finite temperature spin glass phase transition.

Results using neutron scattering \cite{twelve} and Faraday rotation
technique \cite{thirteen} have established random-field induced spin
glass like dynamic behaviour in Fe$_{0.31}$Zn$_{0.69}$F$_{2}$.  Recent
magnetization experiments revealed that a similar behaviour occur at
intense applied fields in samples of Fe$_{x}$Zn$_{1-x}$F$_{2}$, with $x$
= 0.56 and 0.60 \cite{fourteen}. In this paper we discuss experimental
results from dc-magnetisation and ac-susceptibility measurements on
the same Fe$_{0.31}$Zn$_{0.69}$F$_{2}$ system of earlier neutron and
Faraday rotation measurements. In zero applied
field, a slowing down of the dynamics occurs at low temperatures that
obeys a pure Arrhenius law and some slowing down is also observable
near the antiferromagnetic transition temperature. In applied
dc-fields, additional slow dynamical processes are introduced near
$T_{N}$ by the random fields.  A comprehensive static and dynamic
phase diagram in the $H-T$ plane is deduced that, in parts, adequately
compares with an earlier published phase diagram on the same compound
\cite{thirteen}.

\section{experimental}

A high quality single crystal \cite{twelve,thirteen} of
Fe$_{0.31}$Zn$_{0.69}$F$_{2}$ in the form of a parallelepiped with its
longest axis aligned with the crystalline $c$-axis was used as a
sample.  The frequency dependence of the ac-susceptibility in zero
applied dc-field was studied in a Cryogenic Ltd. S600X SQUID-magnetometer.
A commercial Lake Shore 7225 ac-susceptometer was employed for the
ac-susceptibility measurements in a superposed dc magnetic field and
the temperature dependence of the magnetisation in different applied
dc-fields was measured in a Quantum Design MPMS5 SQUID-magnetometer.
The magnetic field was in all experiments applied parallel to the
$c$-axis of the sample.

\section{Results and Discussion}

Fig.  1 shows the temperature dependence of both components of the
ac-susceptibility, (a) $\chi'$($\omega,T$) and (b)

\begin{figure}

\centerline{\hbox{\epsfig{figure=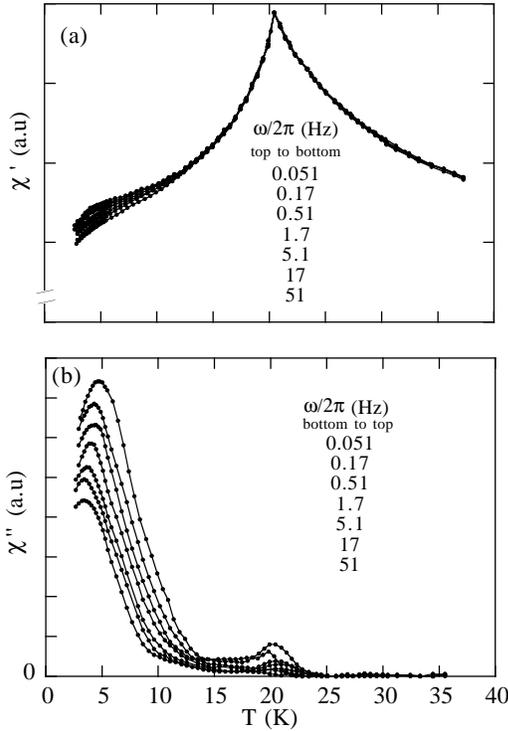,width=7.0 cm}}}
\caption{
\hbox {Temperature dependence of the ac-susceptibility at}
different frequencies as indicated in the figures.  The probing ac field is 1
Oe.  (a) $\chi'(\omega)$ and (b) $\chi''(\omega)$.
}
\label{fig1}
\end{figure}

$\chi''$($\omega,T$).  The different frequencies ranges from 0.051-51
Hz as indicated in the figures.  The transition from a paramagnetic
Curie-Weiss behaviour at high temperatures to long range
antiferromagnetic order is signaled by the cusp in $\chi'$($\omega,T$)
at about 20 K. A small bump in $\chi''$($\omega,T$) is observed at
about the same temperature.  Below 15 K the ac-susceptibility becomes
frequency dependent.  The out-of-phase component increases and a
frequency dependent maximum that shifts towards lower temperatures
with decreasing frequency is observed below $T\approx$ 5 K. The
frequency dependence of $\chi'$($\omega,T$) and $\chi''$($\omega,T$)
at low temperatures shows some resemblance with the behaviour of an
ordinary spin glass.  However, earlier neutron scattering measurements
indicated that AF LRO is established below $T_{N}$ $\approx$
19.8 K in this system \cite {twelve}, provided the sample is submitted to a
slow cooling process. To investigate the nature of the slowing down of the
dynamics
at low temperatures, a comparison is made with the behavior observed in
ordinary spin glasses. A 3d spin glass exhibits conventional
critical slowing down of the dynamics \cite{fifteen} according to:

\begin{equation}
{\frac{\tau }{\tau_{0} } } =      \left({\frac{T_{f}-T_{g} }{T_{g} } }
	\right)^{-z\nu},
\label{conv}
\end{equation}

where $\tau_{0}$ is the microscopic spin flip time of the order
$10^{-13}$-$10^{-14}$ s, $T_{g}$ the spin glass temperature and $z\nu$
a dynamical critical exponent.  Defining the inflection point in
$\chi''$($\omega,T$) as a measure of the freezing temperature $T_{f}$
for a relaxation

\begin{figure}

\centerline{\hbox{\epsfig{figure=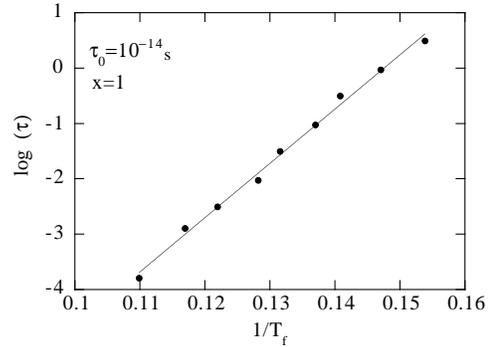,width=6.5cm}}}
\caption{
\hbox {The best fit of the relaxation times to activated} dynamics:
log$t$ vs. $T_{f}^{-1}$, implying a pure Arrhenius behaviour of the
slowing down of the dynamics.
}
\label{fig2}
\end{figure}

time ($\tau$) corresponding to the observation time,
$t\approx$1/$\omega$, of the ac-susceptibility measurement, the
derived data may be employed for dynamic scaling analyses.  The data
do not fit conventional critical slowing down according to eq. 1 with
physically
plausible values of the parameters.  Activated dynamics could
govern the dynamics still yielding a finite phase transition
temperature.  The slowing down of the relaxation times should then
obey:

\begin{equation}
ln\left({\frac{\tau }{\tau_{0} } }\right) = {\frac{1}{T_{f} } }
\left({\frac{T_{f}-T_{g} }{T_{g} } }
	\right)^{-\psi\nu},
\label{activated}
\end{equation}

where $\psi\nu$ is a critical exponent \cite{fh}. The derived data
fits eq. 2 with $T_{g}\approx 0$ which implies that the slowing
down rather is described by a generalized Arrhenius law:

\begin{equation}
log\left({\frac{\tau }{\tau_{0} } }\right) \propto T_{f}^{-x}.
\label{arrhenius}
\end{equation}

Fig. 2 shows the best fit to this expression yielding x=1 and
$\tau_{0}$=10$^{-14}$ s for 0.051 $ \leq \omega/2\pi$(Hz) $\leq$
1000.

The observed frequency dependent ac-susceptibility shows striking
similarities with the behaviour of alleged reentrant antiferromagnets.
In such a system there is a transition from a paramagnetic phase to an
antiferromagnetic phase and spin glass behaviour is observed at low
temperatures.  The reentrant Ising antiferromagnet
Fe$_{0.35}$Mn$_{0.65}$TiO$_{3}$ displays similar features as this
system, e.g.  the low temperature slowing down of the dynamics is
found to obey a pure Arrhenius behaviour \cite{reanti}.

Furthermore,
the more diluted system Fe$_{0.25}$Zn$_{0.75}$F$_{2}$ (on the
percolation threshold) does not display long range antiferromagnetic
order but it exhibits a slowing down of the relaxation times that
follows a pure Arrhenius law \cite {eleven} with a similar value of
$\tau_{0}$ as
here derived for Fe$_{0.31}$Zn$_{0.69}$F$_{2}$.

In Fig. 3 (a) $\chi'$($\omega, T, H$) and (b) $\chi''$($\omega,
T, H$) are plotted for $\omega/2\pi$=125 Hz in different superposed dc
magnetic fields $H\leq$ 2 T.  At these rather low fields, the maximum
in $\chi'$($\omega, T, H$) near $T_{N}$($H$) gets rounded and is
pushed towards lower temperature with increasing magnitude of

\begin{figure}

\centerline{\hbox{\epsfig{figure=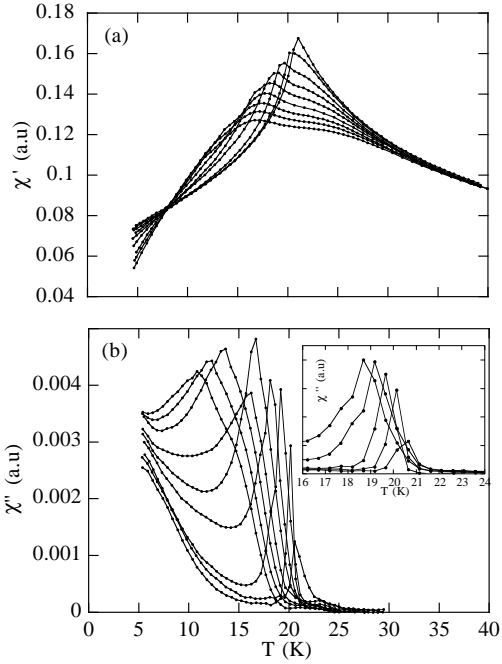,width=8.0cm}}}
\caption{
\hbox {(a) $\chi'(\omega, T, H)$ and (b) $\chi''(\omega, T, H)$ at}
$\omega/2\pi$=125 Hz vs. $T$ in different applied dc-fields: 0, 0.25,
0.5, 0.75, 1, 1.25, 1.5, 1.75 and 2 T. The inset in (b) shows
$\chi''(\omega, T, H)$ for dc-fields 0.25, 0.375, 0.5, 0.625 T. The
amplitude of the ac-field is 10 Oe.
}
\label{fig3}
\end{figure}

the
magnetic field.  The corresponding bump in the out-of-phase component
in zero dc-field, increases in magnitude and sharpens for increasing
dc-fields up to 1 T (the inset of Fig.  3 (b) displays fields up to
0.625 T).  A measure of the phase transition temperature
$T_{N}$($H$) is given by the position of the maximum in the derivative
d($\chi'$$T$)/d$T$ \cite {Fisher relation}. For fields $H \leq$
1.5 T, $T_{N}$($H$) is pushed to lower temperatures with increasing
field strength following a REIM to RFIM crossover scaling, as described in
ref. 13.  At higher fields the maximum is washed out which signals that
the antiferromagnetic phase is destroyed. The destruction of the
antiferromagnetic phase by strong random fields in Fe$_{x}$Zn$_{1-x}$F$_{2}$
was observed by earlier Faraday rotation \cite {thirteen} and neutron
scattering
 \cite {twelve} measurements in the same system ($x$ = 0.31), and by recent
 magnetization \cite {fourteen} and dynamic susceptibility studies
  \cite {sixteen} in less diluted samples ($x$ = 0.42, 0.56 and 0.60).
  A glassy dynamics is found in the upper portion of the $H-T$ phase diagram
  of Fe$_{x}$Zn$_{1-x}$F$_{2}$, at least within the interval $0.31 \leq $x$
\leq 0.60$.

In increasing applied dc-fields the out-of-phase component is enhanced
in a rather narrow but widening region near the antiferromagnetic
phase transition due to the introduction of random fields that create new
slow dynamical processes in the system.  The increase of
$\chi''$($\omega, T, H$) at lower temperatures, corresponding to the
processes causing the slowing down of the dynamics already in zero
field, remains observable also when the field is increased. This latter feature
cannot be entirely attributed

\begin{figure}

\centerline{\hbox{\epsfig{figure=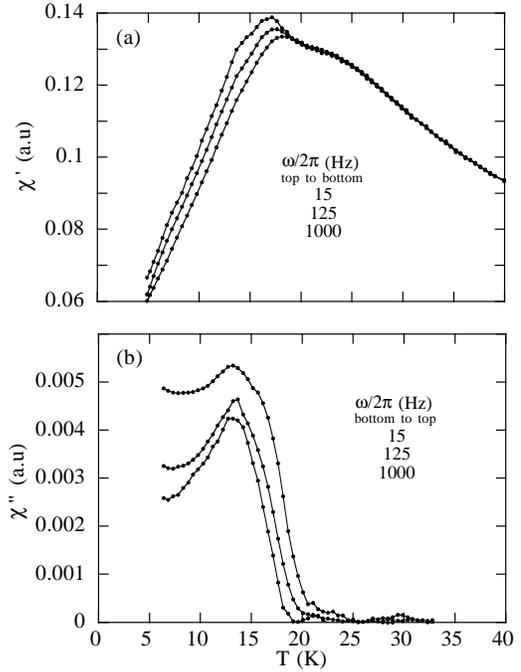,width=7.0cm}}}
\caption{
\hbox {(a) $\chi'(\omega, T,
H)$ and (b) $\chi''(\omega, T, H)$ vs.
$T$}
in an applied dc-field $H$=1.5 T and at different frequencies
$\omega/2\pi$=15, 125 and 1000 Hz. The
amplitude of the ac-field is 10 Oe.
}
\label{fig4}
\end{figure}

to random fields. For larger
fields these low temperature processes and the processes caused by the
random fields start to overlap, and at the highest dc-fields they even
become indistinguishable.  In Fig.  4 both components of the
ac susceptibility are plotted, in an applied dc-field $H$=1.5 T,
for $\omega/2\pi$=15, 125 and 1000 Hz. Note that the temperature
of the maximum in $\chi'$($\omega, T, H$), at $T_{N} (H)$, shifts to
lower temperatures as the frequency decreases. By way of contrast,
no shift in the peak temperature is observable as a function of the
frequency in dynamic susceptibility measurements performed
in Fe$_{0.46}$Zn$_{0.54}$F$_{2}$ \cite {test}
and Fe$_{0.42}$Zn$_{0.58}$F$_{2}$ \cite {sixteen}, within the field
limits of the weak RFIM problem in each case. The frequency dependent
behaviour of $T_{N} (H)$ is a feature associated with the effects of strong
random fields in samples of Fe$_{x}$Zn$_{1-x}$F$_{2}$,
particularly with $x$ close to $x_p$.

In Fig. 5 (a) and (b) $\chi'$($\omega, T, H$) and $\chi''$($\omega, T,
H$) are plotted for $\omega/2\pi$=125 Hz in different superposed dc
magnetic fields $H\geq$2 T. The maximum in the in-phase-component is
flattend, the susceptibility is strongly surpressed and the onset of the
out-of-phase susceptibility is shifted towards lower temperatures as
the dc-field is increased.  No sign of a transition to an antiferromagnetic
phase is observed.

Fig.  6 shows the temperature dependence of the field cooled (FC),
$M_{FC}$($T$)/$H$, and zero field cooled (ZFC), $M_{ZFC}$($T$)/$H$,
susceptibility \cite{seventeen} at three different applied
magnetic
fields.  Below a temperature $T_{ir}$ the magnetisation becomes
irreversible.  $T_{ir}$ decreases with increasing

\begin{figure}

\centerline{\hbox{\epsfig{figure=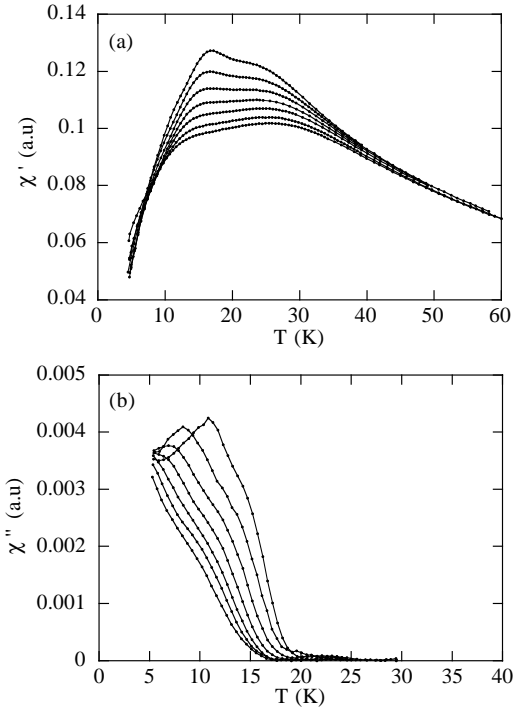,width=7.0cm}}}
\caption{
\hbox {(a) $\chi'(\omega, T, H)$ and (b) $\chi''(\omega, T, H)$ at}
$\omega/2\pi$=125 Hz vs. $T$ in different applied dc-fields: 2, 2.5,
3, 3.5, 4, 4.5 and 5 T. The
amplitude of the ac-field is 10 Oe.
}
\label{fig5}
\end{figure}

\begin{figure}

\centerline{\hbox{\epsfig{figure=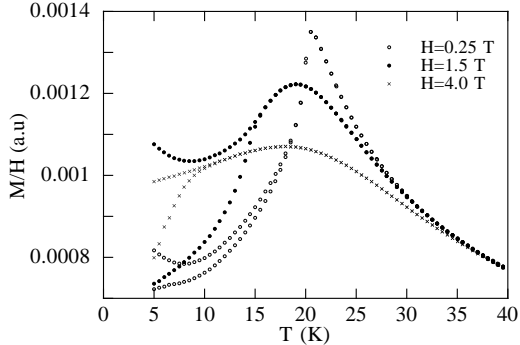,width=7.0cm}}}
\caption{
\hbox {Temperature dependence of the dc-susceptibility} for
zero-field-cooled (ZFC) and field-cooled (FC) procedures in different
fields, as indicated in the figure.
}
\label{fig6}
\end{figure}

 field.  The
irreversibility point is associated with an observation time mainly
governed by the heating rate of the ZFC experiment which in our
experiment corresponds to about 100 s.

In Fig. 7 an $H-T$ magnetic phase diagram is shown,
in which some of the
above discussed experimental characteristics are summarised.  The open circles
represent $T_{N}$($H$), the solid circles $T_{ir}$($H$), diamonds
the spin freezing temperatures $T_{f}$($H$) for $\omega
/2\pi$=125 Hz and open triangels label $T_{f}$($H$=0) for different
frequencies. The onset of  $\chi''$($\omega, T, H$) at frequencies
$\omega/2\pi$=15, 125 and 1000 Hz are shown as solid triangels, solid
squares and open squares respectively. Those are measures that
mirror the observation time dependence of $T_{ir}$.

\begin{figure}

\centerline{\hbox{\epsfig{figure=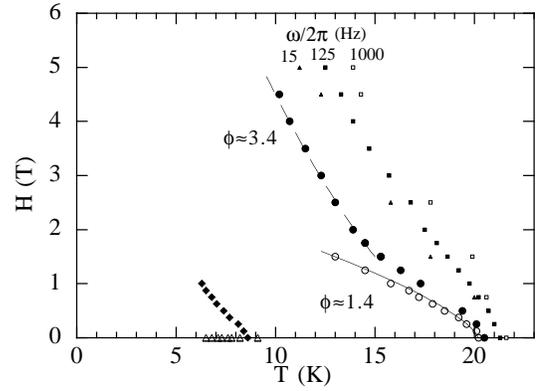,width=7.0 cm}}}
\caption{
\hbox {$H-T$ diagram of the Fe$_{0.31}$Zn$_{0.69}$F$_{2}$ }system.
$T_{N}$($H$) open circles, $T_{ir}$($H$) solid circles, onset of
  $\chi''(\omega, T, H)$ for $\omega/2\pi$=15 (solid triangels), 125
  (solid squares) and 1000 Hz (open squares),
  $T_{f}$($H$, $\omega/2\pi$=125 Hz) diamonds and $T_{f}$($H$=0)
 triangels for different frequencies $\omega/2\pi$ (from left to
 right): 0.051, 0.17, 0.51, 1.7, 5.1, 17, 51, 125 and 1000 Hz.
}
\label{fig7}
\end{figure}

In diluted Ising antiferromagnets, $T_{N}$ is expected to decrease with
increasing
magnetic fields as:

\begin{equation}
{\epsilon  \propto      H
	^{2/\phi} \qquad \textnormal{;}\qquad
\epsilon=\left({\frac{T_{N}(H)-T_{N}(0)+bH^{2}
	}{T_{N}(0) } }
	\right)
}
\label{less}
\end{equation}

where $\phi$ is a crossover exponent and $bH^2$ a small mean field
correction.  For low fields, $H\leq$1.5 T, we find $\phi\approx$1.4
using $b$=0 for $T_{N}$($H$) as indicated by the solid line in Fig.
7.  For higher fields, $H\geq$1.5 T, a reversal of the curvature of
$T_{ir}$($H$) occurs.  The dashed line corresponds to a functional
behaviour according to eq. 4 with an exponenet $\phi\approx$ 3.4.  A
largely equivalent phase diagram has earlier been established for the
same system utilising Faraday rotation measurements \cite{thirteen}.
One significant difference being that $T_{ir}$($0$) $\approx
T_{N}$($0$) in ref.  \cite{thirteen}, whereas we find a significant
difference between these two temperatures, as is also observed in other dilute
antiferromagnets \cite{voffe}.  The field dependence of $T_{N}$($H$)
is equivalent to those of the more concentrated
Fe$_{0.46}$Zn$_{0.54}$F$_{2}$ and Fe$_{0.72}$Zn$_{0.28}$F$_{2}$
where the scaling behaviour of eq. 4 gives $\phi \approx$1.4 for fields
up to 2 T and 10 T respectively \cite{eighteen}.  The new features of
the phase diagram in Fig.  7 as compared to the one of ref.
\cite{thirteen} are the observation
time dependent spin freezing
temperatures at low temperature and the observation time dependence of
$T_{ir}$($H$) demonstrated by the shifts of the $T_{ir}$($H$) contours
towards higher temperatures when decreasing the observation time.
A possible mechanism for the spin freezing at low temperatures may be a
weak frustration present in a third nearest neighbour interaction of this
compound. Results of numerical simulation \cite {nineteen} indicates that a
small frustrated bond plays no role in the REIM properties of
Fe$_{x}$Zn$_{1-x}$F$_{2}$ under weak dilution. However, it causes dramatic
influences in the antiferromagnetic and spin glass order parameters close
to the percolation threshold.

\section{conclusions}

Dynamic and static magnetic properties of the diluted antiferromagnet
Fe$_{0.31}$Zn$_{0.69}$F$_{2}$ have been studied.  The dynamic
susceptibility in zero dc-field shows similarities to a reentrant
Ising antiferromagnet with a slowing down of the dynamics at low
temperatures best described by a pure Arrhenius law.  Hence, there is
no transition to a spin glass phase at low temperatures.

The field dependence of the antiferromagnetic
transition temperature follows the predicted scaling behaviour for a
random field system, in accord with earlier experimental
findings\cite{thirteen,eighteen}. The onset of $\chi''(\omega, T, H)$ occur
above the antiferromagnetic phase transition, even in zero
applied magnetic field. $\chi''(\omega, T, H)$ shows a frequency dependent
behaviour that mirror the observation time dependence of the FC-ZFC
irreversibility  line. The dynamics of the diluted antiferromagnet
Fe$_{0.31}$Zn$_{0.69}$F$_{2}$ has been shown to involve not only
random field induced slow dynamics near $T_{N}$($H$), but additional
slow dynamics originating from the strong dilution appears at low
temperatures.

\section{acknowledgments}

Financial support from the Swedish Natural Science Research Council
(NFR) is acknowledged. One of the authors (FCM) acknowledge the support
from CNPq and FINEP (Brazilian agencies).

\end{multicols}


\begin {references}
\bibitem{seventeen} Susceptibility is here defined and calculated as
$M/H$.
\end{references}
\end{document}